\input harvmac
\input graphicx
\input color

\def\Title#1#2{\rightline{#1}\ifx\answ\bigans\nopagenumbers\pageno0\vskip1in
\else\pageno1\vskip.8in\fi \centerline{\titlefont #2}\vskip .5in}

%
%
\ifx\includegraphics\UnDeFiNeD\message{(NO graphicx.tex, FIGURES WILL BE IGNORED)}
\def\figin#1{\vskip2in}
\else\message{(FIGURES WILL BE INCLUDED)}\def\figin#1{#1}
\fi
\def\Fig#1{Fig.~\the\figno\xdef#1{Fig.~\the\figno}\global\advance\figno
 by1}
%
%
%
%

\font\ticp=cmcsc10

\def \purge#1 {\textcolor{magenta}{#1}}
\def \new#1 {\textcolor{blue}{#1}}
\def\comment#1{}

\def\\{\cr}
\def\text#1{{\rm #1}}
\def\frac#1#2{{#1\over#2}}



\def\calh{{\cal H}}

\def\roughly#1{\mathrel{\raise.3ex\hbox{$#1$\kern-.75em\lower1ex\hbox{$\sim$}}}}
\font\bbbi=msbm10 
\def\mathbb#1{\hbox{\bbbi #1}}

\def\mthsu{\mathsurround=0pt  }
\def\leftrightarrowfill{$\mthsu \mathord\leftarrow\mkern-6mu\cleaders
  \hbox{$\mkern-2mu \mathord- \mkern-2mu$}\hfill
  \mkern-6mu\mathord\rightarrow$}
\def\overleftrightarrow#1{\vbox{\ialign{##\crcr\leftrightarrowfill\crcr\noalign{\kern-1pt\nointerlineskip}$\hfil\displaystyle{#1}\hfil$\crcr}}}
\overfullrule=0pt

%
%
\lref\QFG{
  S.~B.~Giddings,
  ``Quantum-first gravity,''
[arXiv:1803.04973 [hep-th]].
}
\lref\MaSu{
  J.~Maldacena and L.~Susskind,
  ``Cool horizons for entangled black holes,''
Fortsch.\ Phys.\  {\bf 61}, 781 (2013).
[arXiv:1306.0533 [hep-th]].
}
\lref\Witt{
  E.~Witten,
  ``Notes on Some Entanglement Properties of Quantum Field Theory,''
[arXiv:1803.04993 [hep-th]].
}
\lref\NVNLT{
  S.~B.~Giddings,
  ``Modulated Hawking radiation and a nonviolent channel for information release,''
[arXiv:1401.5804 [hep-th]].
}
\lref\BHQIUE{
  S.~B.~Giddings,
  ``Black holes, quantum information, and unitary evolution,''
  Phys.\ Rev.\ D {\bf 85}, 124063 (2012).
[arXiv:1201.1037 [hep-th]].
}
\lref\SGmodels{
  S.~B.~Giddings,
   ``Models for unitary black hole disintegration,''  Phys.\ Rev.\ D {\bf 85}, 044038 (2012)
[arXiv:1108.2015 [hep-th]].
}
\lref\NVNL{
  S.~B.~Giddings,
  ``Nonviolent nonlocality,''
  Phys.\ Rev.\ D {\bf 88},  064023 (2013).
[arXiv:1211.7070 [hep-th]].
}
\lref\Locbdt{
  S.~B.~Giddings and M.~Lippert,
  ``The Information paradox and the locality bound,''
Phys.\ Rev.\ D {\bf 69}, 124019 (2004).
[hep-th/0402073].
}
\lref\BCS{
  N.~Bao, S.~M.~Carroll and A.~Singh,
  ``The Hilbert Space of Quantum Gravity Is Locally Finite-Dimensional,''
Int.\ J.\ Mod.\ Phys.\ D {\bf 26}, no. 12, 1743013 (2017).
[arXiv:1704.00066 [hep-th]].
}
\lref\Chru{Piotr T. Chru\'sciel, ``Anti-gravity \`a la Carlotto-Schoen,"  arXiv:1611.01808 [math.DG].}
\lref\DoGithree{
  W.~Donnelly and S.~B.~Giddings,
  ``How is quantum information localized in gravity?,''
Phys.\ Rev.\ D {\bf 96}, no. 8, 086013 (2017).
[arXiv:1706.03104 [hep-th]].
}
\lref\DoGifour{W.~Donnelly and S.~B.~Giddings, unpublished.
}
\lref\UQM{
  S.~B.~Giddings,
  ``Universal quantum mechanics,''
Phys.\ Rev.\ D {\bf 78}, 084004 (2008).
[arXiv:0711.0757 [quant-ph]].
}
\lref\Haag{R. Haag, {\sl Local quantum physics, fields, particles, algebras}, Springer (Berlin, 1996).}
\lref\SGalg{
  S.~B.~Giddings,
 ``Hilbert space structure in quantum gravity: an algebraic perspective,''
JHEP {\bf 1512}, 099 (2015).
[arXiv:1503.08207 [hep-th]].
}
\lref\locbdi{
  S.~B.~Giddings and M.~Lippert,
  ``Precursors, black holes, and a locality bound,''
Phys.\ Rev.\ D {\bf 65}, 024006 (2002).
[hep-th/0103231].
}
\lref\DoGione{
  W.~Donnelly and S.~B.~Giddings,
  ``Diffeomorphism-invariant observables and their nonlocal algebra,''
Phys.\ Rev.\ D {\bf 93}, no. 2, 024030 (2016), Erratum: [Phys.\ Rev.\ D {\bf 94}, no. 2, 029903 (2016)].
[arXiv:1507.07921 [hep-th]].
}
\lref\DoGitwo{
  W.~Donnelly and S.~B.~Giddings,
  ``Observables, gravitational dressing, and obstructions to locality and subsystems,''
Phys.\ Rev.\ D {\bf 94}, no. 10, 104038 (2016).
[arXiv:1607.01025 [hep-th]].
}
\lref\HPS{
  S.~W.~Hawking, M.~J.~Perry and A.~Strominger,
  ``Soft Hair on Black Holes,''
Phys.\ Rev.\ Lett.\  {\bf 116}, no. 23, 231301 (2016).
[arXiv:1601.00921 [hep-th]]\semi
``Superrotation Charge and Supertranslation Hair on Black Holes,''
JHEP {\bf 1705}, 161 (2017).
[arXiv:1611.09175 [hep-th]].
}
\lref\CCM{
  C.~Cao, S.~M.~Carroll and S.~Michalakis,
  ``Space from Hilbert Space: Recovering Geometry from Bulk Entanglement,''
Phys.\ Rev.\ D {\bf 95}, no. 2, 024031 (2017).
[arXiv:1606.08444 [hep-th]].
}
\lref\CaCa{
  C.~Cao and S.~M.~Carroll,
  ``Bulk Entanglement Gravity without a Boundary: Towards Finding Einstein's Equation in Hilbert Space,''
[arXiv:1712.02803 [hep-th]].
}
\lref\CaSi{
  S.~M.~Carroll and A.~Singh,
  ``Mad-Dog Everettianism: Quantum Mechanics at Its Most Minimal,''
[arXiv:1801.08132 [quant-ph]].
}
\lref\Hartone{
  J.~B.~Hartle,
  ``The Quantum mechanics of cosmology,''
  in {\sl Quantum cosmology and baby universes : proceedings}, 7th Jerusalem Winter School for Theoretical Physics, Jerusalem, Israel, December 1989, ed. S. Coleman, J. Hartle, T. Piran, and S. Weinberg (World Scientific, 1991).
}
\lref\Harttwo{
  J.~B.~Hartle,
  ``Space-time coarse grainings in nonrelativistic quantum mechanics,''
  Phys.\ Rev.\  D {\bf 44}, 3173 (1991).
}
\lref\HartLH{
  J.~B.~Hartle,
  ``Space-Time Quantum Mechanics And The Quantum Mechanics Of Space-Time,''
  arXiv:gr-qc/9304006.
}
\lref\HartPuri{
  J.~B.~Hartle,
  ``Quantum Mechanics At The Planck Scale,''
  arXiv:gr-qc/9508023.
}
\lref\Eins{A. Einstein, ``Quanten-Mechanik und Wirklichkeit," Dialectica {\bf 2} (1948) 320.}
\lref\Howa{For translation, see D. Howard, ``Einstein on locality and separability," Studies in History and Philosophy of Science Part A  {\bf 16} no. 3, (1985)
171.}
\lref\vanR{
  M.~Van Raamsdonk,
 ``Building up spacetime with quantum entanglement,''
Gen.\ Rel.\ Grav.\  {\bf 42}, 2323 (2010), [Int.\ J.\ Mod.\ Phys.\ D {\bf 19}, 2429 (2010)].
[arXiv:1005.3035 [hep-th]].
}
\lref\GiRo{
  S.~B.~Giddings and M.~Rota,
 ``Quantum information/entanglement transfer rates between subsystems,''
[arXiv:1710.00005 [quant-ph]].
}
\lref\BRSSZ{
  A.~R.~Brown, D.~A.~Roberts, L.~Susskind, B.~Swingle and Y.~Zhao,
  ``Complexity, action, and black holes,''
Phys.\ Rev.\ D {\bf 93}, no. 8, 086006 (2016).
[arXiv:1512.04993 [hep-th]].
}
\lref\Sussfall{
  L.~Susskind,
  ``Why do Things Fall?,''
[arXiv:1802.01198 [hep-th]].
}
\lref\NPGNL{
  S.~B.~Giddings,
  ``(Non)perturbative gravity, nonlocality, and nice slices,''
Phys.\ Rev.\ D {\bf 74}, 106009 (2006).
[hep-th/0606146].
}
\lref\Mald{
  J.~M.~Maldacena,
  ``Eternal black holes in anti-de Sitter,''
JHEP {\bf 0304}, 021 (2003).
[hep-th/0106112].
}
\lref\Heem{
  I.~Heemskerk,
  ``Construction of Bulk Fields with Gauge Redundancy,''
JHEP {\bf 1209}, 106 (2012).
[arXiv:1201.3666 [hep-th]].
}
\lref\KaLigrav{
  D.~Kabat and G.~Lifschytz,
  ``Decoding the hologram: Scalar fields interacting with gravity,''
Phys.\ Rev.\ D {\bf 89}, no. 6, 066010 (2014).
[arXiv:1311.3020 [hep-th]].
}
\lref\Torr{
  C.~G.~Torre,
 ``Gravitational observables and local symmetries,''
Phys.\ Rev.\ D {\bf 48}, R2373 (1993).
[gr-qc/9306030].
}
\lref\Zure{W. H. Zurek, ``Quantum darwinism, classical reality, and the 
randomness of quantum jumps," Physics Today {\bf 67} 44, 	arXiv:1412.5206.
}
\lref\LQGST{
  S.~B.~Giddings,
 ``Locality in quantum gravity and string theory,''
Phys.\ Rev.\ D {\bf 74}, 106006 (2006).
[hep-th/0604072].
}
\lref\GiKi{
  S.~B.~Giddings and A.~Kinsella,
  ``Gauge-invariant observables, gravitational dressings, and holography in AdS,''
[arXiv:1802.01602 [hep-th]].
}
\lref\BuVe{
  D.~Buchholz and R.~Verch,
  ``Scaling algebras and renormalization group in algebraic quantum field theory,''
Rev.\ Math.\ Phys.\  {\bf 7}, 1195 (1995).
[hep-th/9501063].
}
\lref\HSTone{
  T.~Banks and W.~Fischler,
  ``M theory observables for cosmological space-times,''
[hep-th/0102077].
}
\lref\Yngv{
  J.~Yngvason,
  ``The Role of type III factors in quantum field theory,''
Rept.\ Math.\ Phys.\  {\bf 55}, 135 (2005).
[math-ph/0411058].
}
\lref\CaSc{A Carlotto and R Schoen, ``Localizing solutions of the Einstein constraint equations," Invent.
Math. {\bf 205}, 559 (2016).}
\lref\HSTrev{
  T.~Banks,
  ``Lectures on Holographic Space Time,''
[arXiv:1311.0755 [hep-th]].
}
\lref\Hawkevap{
  S.~W.~Hawking,
  ``Black hole explosions,''
Nature {\bf 248}, 30 (1974)\semi
``Particle Creation by Black Holes,''
Commun.\ Math.\ Phys.\  {\bf 43}, 199 (1975), Erratum: [Commun.\ Math.\ Phys.\  {\bf 46}, 206 (1976)].
}
\lref\Hawkunc{
  S.~W.~Hawking,
  ``Breakdown of Predictability in Gravitational Collapse,''
Phys.\ Rev.\ D {\bf 14}, 2460 (1976).
}
\lref\ZLL{
  P.~Zanardi, D.~A.~Lidar and S.~Lloyd,
 ``Quantum tensor product structures are observable induced,''
Phys.\ Rev.\ Lett.\  {\bf 92}, 060402 (2004).
[quant-ph/0308043].
}
\lref\Harl{
  D.~Harlow,
  ``Wormholes, Emergent Gauge Fields, and the Weak Gravity Conjecture,''
JHEP {\bf 1601}, 122 (2016).
[arXiv:1510.07911 [hep-th]].
}
\lref\DoFr{
  W.~Donnelly and L.~Freidel,
  ``Local subsystems in gauge theory and gravity,''
JHEP {\bf 1609}, 102 (2016).
[arXiv:1601.04744 [hep-th]].
}
\lref\GuJa{
  M.~Guica and D.~L.~Jafferis,
  ``On the construction of charged operators inside an eternal black hole,''
SciPost Phys.\  {\bf 3}, no. 2, 016 (2017).
[arXiv:1511.05627 [hep-th]].
}
\lref\CPR{
  J.~S.~Cotler, G.~R.~Penington and D.~H.~Ranard,
  ``Locality from the Spectrum,''
[arXiv:1702.06142 [quant-ph]].
}
\lref\PaRa{
  K.~Papadodimas and S.~Raju,
  ``Local Operators in the Eternal Black Hole,''
Phys.\ Rev.\ Lett.\  {\bf 115}, no. 21, 211601 (2015).
[arXiv:1502.06692 [hep-th]].
}
\lref\CoSc{
  J.~Corvino and R.~M.~Schoen,
  ``On the asymptotics for the vacuum Einstein constraint equations,''
J.\ Diff.\ Geom.\  {\bf 73}, no. 2, 185 (2006).
[gr-qc/0301071].
}
\lref\ChDe{
  P.~T.~Chru\'sciel and E.~Delay,
  ``On mapping properties of the general relativistic constraints operator in weighted function spaces, with applications,''
Mem.\ Soc.\ Math.\ France {\bf 94}, 1 (2003).
[gr-qc/0301073].
}
\lref\NVNLpost{
  S.~B.~Giddings,
  ``Nonviolent unitarization: basic postulates to soft quantum structure of black holes,''
JHEP {\bf 1712}, 047 (2017).
[arXiv:1701.08765 [hep-th]].
}
\Title{
\vbox{\baselineskip12pt  
}}
{\vbox{\centerline{Quantum gravity: a quantum-first approach$^\dagger$
} }}
\footnote{}{${}^\dagger$ Abridged and rewritten version of \refs{\QFG}; essay written for the Gravity Research Foundation 2018 Awards for Essays on Gravitation.}

\centerline{{\ticp 
Steven B. Giddings\footnote{$^\ast$}{Email address: giddings@ucsb.edu}
} }
\centerline{\sl Department of Physics}
\centerline{\sl University of California}
\centerline{\sl Santa Barbara, CA 93106}
\vskip.10in
\centerline{\bf Abstract}
A ``quantum-first" approach to gravity is described, where rather than quantizing general relativity, one 
seeks to formulate the physics of gravity within a quantum-mechanical framework with suitably general postulates. Important guides are the need for appropriate mathematical structure on Hilbert space, and correspondence with general relativity and quantum field theory in weak-gravity situations.  A basic physical question is that of ``Einstein separability:" how to define mutually independent subsystems, {\it e.g.} through localization.  Standard answers via tensor products or operator algebras conflict with properties of gravity, as is seen in the correspondence limit; this connects with discussions of ``soft hair."  Instead, 
gravitational behavior suggests a networked Hilbert space structure.  This structure plus unitarity provide important clues towards a quantum formulation of gravity.

\Date{March 31, 2018}

The most profound foundational problem in physics is probably that of reconciling gravity with the reality of the quantum world.  Attempts to {\it quantize} gravity, beginning with general relativity (GR) and applying certain rules (canonical quantization, functional integration, {\it etc.}) have met  vexing difficulties: nonrenormalizability, and the deeper problem of nonunitarity, associated with black hole decay\refs{\Hawkevap,\Hawkunc}.  String theory follows a similar path, and despite addressing nonrenormalizability, still confronts nonunitarity.  But quantum mechanics is quite rigid, and that suggests another approach:  begin with quantum mechanics (QM), rather than with spacetime, and within QM find the structure needed to describe gravity.

Conceivably gravity requires modification of QM, but we'll explore whether QM suffices.  However, in typical formulations QM has extra elements likely not present in quantum gravity; we need to begin with a suitably general framework.  One such approach was Hartle's ``generalized QM~\refs{\Hartone\Harttwo\HartLH-\HartPuri}," but that is still tied to the notion of quantizing spacetime.  More basically,  essential quantum principles seem to be the existence of a  linear space of states with inner product (``Hilbert space"), hermitian operators interpreted as quantum observables, and in appropriate contexts unitarity, {\it e.g.} of the S-matrix.  These  ``universal QM\UQM" principles are notably sparse.

While these principles distill the spacetime-independent content of QM, clearly more mathematical structure is needed on Hilbert space to describe gravity.  A key question was enunciated by Einstein\refs{\Eins,\Howa}: ``... it appears to be essential for this arrangement of the things introduced in physics that, at a specific time, these things claim an existence independent of one another, insofar as these things `lie in different parts of space.'  Without such an assumption of the mutually independent existence (the `being-thus') of spatially distant things, an assumption which originates in everyday thought, physical thought in the sense familiar to us would not be possible.  Nor does one see how physical laws could be formulated and tested without such a clean separation."

As a starting point for a mathematically consistent structure to build on, we thus look for a notion of ``subsystems," providing such separability.  Subsystem structure is hardwired into other quantum theories.  In lattice systems, subsystems arise from tensor factors of the Hilbert space.  Local quantum field theory (LQFT) is more subtle, due to ``infinite entanglement" between neighboring regions (the type-III property of its von Neumann algebras); instead one defines subsystems using commuting subalgebras of operators associated to spacelike-separated regions\Haag\foot{These issues are nicely reviewed in \refs{\Witt}.} -- matching Einstein's description. But the question is what provides the correct underlying mathematical structure, or ``gravitational substrate," on a Hilbert space $\calh$ with gravity.

For a guide beyond mathematical consistency, we appeal to {\it correspondence}:  in weak gravity regimes with small perturbations about semiclassical spacetimes, we assume the fundamental theory must 
approximately match LQFT plus perturbative GR, working in an expansion in the gravitational coupling $\kappa=\sqrt{32\pi G}$.  There is abundant evidence for this in experimental physics.

So our ``quantum-first" approach is to begin with the essential principles of QM, and then follow the  twin guides of need for a mathematically consistent structure and of correspondence.  This minimalist approach assumes no extra hidden degrees of freedom, or extraneous structure.  These guides are actually quite nontrivial.

A similar approach to LQFT provides an example.  LQFT can be viewed as a solution to the problem of constructing a quantum theory, implementing the additional postulates of spacetime locality and special relativity.  Locality is introduced through the algebraic structure\refs{\Haag,\Witt} described above: one has privileged subalgebras of observables, associated with the open sets of spacetime, which commute at spacelike separation.  In the limit of small neighborhoods, these give local fields.  This extra structure thus arises directly from the underlying spacetime manifold. Relativity is implemented through an action of the Poincar\'e group.  

We  seek  analogous structure for gravity by investigating its weak-field correspondence limit.   Spacetime is used to infer the perturbative structure of the theory, but ultimately we seek to determine the intrinsic mathematical structure on the Hilbert space, in which spacetime is not expected to be fundamental.
In parallel with LQFT, consider properties of quantum observables, for concreteness in the theory of a scalar $\phi$ coupled to gravity.  In gravity, $\phi(x)$ is no longer a gauge-invariant observable: diffeomorphisms act via
\eqn\diffdef{\delta x^\mu=-\kappa \xi^\mu(x)\ ,}
so act nontrivially on local operators.  One can find gauge-invariants by ``dressing\refs{\Heem\KaLigrav-\DoGione}" $\phi(x)$, which at linear order in $\kappa$ becomes
\eqn\dressop{\Phi(x)=\phi(x^\mu+V^\mu(x))\ .}
The $V^\mu(x)$ are written in terms of the metric perturbation,
\eqn\metpert{g_{\mu\nu}(x)=\eta_{\mu\nu}+\kappa h_{\mu\nu}(x)}
and are constrained by diffeomorphism invariance, using
\eqn\hdiff{\delta h_{\mu\nu}= -\partial_\mu\xi_\nu - \partial_\nu \xi_\mu\ ,}
or equivalently by the condition that $\Phi(x)$ commutes with the constraints $G_0^\mu-8\pi GT^\mu_0=0$. Different choices exist\DoGione. One is defined in terms of integrals along an arbitrary
curve $\Gamma$ connecting $(x,\infty)$,
\eqn\Vline{ V_\mu^\Gamma(x)= {\kappa\over 2} \int_x^\infty dx^{\prime\nu} \left\{ h_{\mu\nu}(x') + \int_{x'}^\infty dx^{\prime\prime\lambda}\left[\partial_\mu h_{\nu\lambda}(x'') - \partial_\nu h_{\mu\lambda}(x'')\right]\right\}\ ;}
then $\Phi(x)$ creates a particle together with a gravitational field concentrated along $\Gamma$. Another\DoGione, $V_\mu^C(x)$, found from an average of \Vline\ over directions, creates a linearized Schwarzschild field.

We ask what are  {\it intrinsic} properties of the operators that should match the  fundamental theory\SGalg.
One finds\DoGione
\eqn\commfail{[\Phi(x),\Phi(y)]\neq0}
for spacelike $x-y$, due to the dressing $V_\mu(x)$ extending to infinity; gravity has no local commuting observables\refs{\Torr,\DoGitwo}.

With neither tensor factorization nor commuting subalgebras to define subsystems and separability we need another approach.  Is there some notion of ``independent information" in a region, not evident in the gravitational field outside, or put differently, ``what is a localized qubit in gravity?"  Part of the answer is provided in \refs{\DoGithree,\DoGifour}:  given a matter configuration in a compact spatial region (extendable to a spacetime region), one may always find a linearized gravitational field outside that depends only on the total Poincar\'e charges of the matter.  This field may be linelike as from \Vline\ -- incidentally providing an example of local screening of gravity, or ``antigravity\refs{\CaSc,\Chru}" -- or linearized boosted Kerr.\foot{Generalizing\DoGithree\ the Corvino-Schoen gluing theorem\refs{\CoSc,\ChDe}.}

Different localized information may thus be encoded in different configurations with identical Poincar\'e charges and thus asymptotic fields.
More generally, this suggests the notion of a ``gravitational split structure\DoGithree" associated to a neighborhood $U$, which is a Hilbert subspace of states $\calh^i_U\subset \calh$ so that for two states $|\psi \rangle,\ |\tilde \psi \rangle\in \calh^i_{U}$ and any observable $A$ localized outside $U$, 
\eqn\splstr{ \langle \tilde \psi|  A|\psi \rangle = \langle \tilde \psi| \psi \rangle \langle i|A| i\rangle\ :}
 the ``value" of $A$ only depends on the Hilbert space label $i$.

A suggested alternative to the algebraic structure of LQFT is thus a network of such Hilbert subspaces related by inclusion maps.  For example, if $U'\subset U$, we expect an inclusion
\eqn\subinc{\calh^i_{U'}\hookrightarrow \calh^i_{U}\ .}
{\it However}, there should not be an  $\calh^i_U\subset \calh$ for {\it each} $U$; if we try to excite a state in a neighborhood small compared to the Planck length $\sim\kappa$, its strong gravitational field extends outside the neighborhood.  Similarly, for a separated pair $U,U'$ we expect
\eqn\prodinc{\calh_U^i\otimes \calh_{U'}^{i'} \hookrightarrow \calh\ ,}
with limitations for states producing a strong gravitational field spanning the separation.  These limitations are in fact seen in \commfail: noncommutativity becomes large for such strong fields\refs{\locbdi\Locbdt-\LQGST,\DoGione}.

Such a network of Hilbert space inclusions provides a candidate substrate.  While spacetime has been used to infer the perturbative limit of the mathematical structure, this network - once more completely determined -- is more basic, providing an example of a possible quantum replacement for spacetime.

Similar approaches have begun with tensor products of Hilbert spaces\refs{\HSTone\HSTrev\BHQIUE\CCM\CaCa-\CaSi}, but we have found 
gravitational systems don't have Hilbert spaces with simple tensor factorizations.
Another approach, ``spacetime from entanglement\refs{\vanR,\MaSu}," begins with tensor product structure, and is puzzling  from the present view, which has argued that definition of a subsystem structure {\it arises from} an analog of spacetime structure, and is {\it prior} to definition of entanglement\refs{\QFG}; moreover, we expect this subsystem structure to be hardwired into  locality properties of the hamiltonian.

The mathematical structure implementing Einstein's separability is expected to play a key structural role and constrain the theory.  A related question is that of evolution, particularly given the problems black holes present for unitarity.  Is a black hole (BH) a localized subsystem?  In Hawking's original LQFT-based argument\refs{\Hawkevap,\Hawkunc} it is, and its ultimate disappearance spoils unitary evolution.  

But we have found this question is more subtle; a related ``soft quantum hair" proposal\refs{\HPS} has even suggested that gravitational delocalization of information might restore unitarity.  This is an important question.  The preceding arguments regarding split structure suggest that gravitational dressings may be found that make the ``soft charges" of \refs{\HPS} corresponding to a given matter configuration vanish, so the soft hair doesn't encode its information, though higher-order questions remain\refs{\QFG}.  

If indeed the information in states of a BH subsystem is invisible from outside, unitary evolution apparently requires transfer of this information out of the BH, through new interactions not described by LQFT.  One can attempt to parameterize the needed interactions, in a form not spoiling correspondence\refs{\SGmodels,\BHQIUE,\NVNL,\NVNLpost}.  If the subsystem structure requires such interactions to be present to preserve unitarity, that is also very important information about the  theory.

A quantum-first approach thus provides a tight framework in which to cast gravity.  The twin goals of finding mathematically consistent structure needed for physics, and respecting correspondence with quantum field theory and perturbative general relativity, provide extremely constraining guides.  We already find  quite nontrivial mathematical structure in the perturbative/correspondence limit.  Plausibly these combined requirements, including unitarity, will give key clues to a theory explaining gravity within quantum mechanics.

\bigskip\bigskip\centerline{{\bf Acknowledgments}}\nobreak

This material is based upon work supported in part by the U.S. Department of Energy, Office of Science, under Award Number {DE-SC}0011702.

\listrefs
\end